**Commensurability between element symmetry and the number of skyrmions governing skyrmion diffusion in confined geometries**

*Chengkun Song, Nico Kerber, Jan Rothörl, Yuqing Ge, Klaus Raab, Boris Seng, Maarten A. Brems, Florian Dittrich, Robert M. Reeve, Jianbo Wang, Qingfang Liu, Peter Virnau and Mathias Kläui* [*]

Chengkun Song, Prof. Jianbo Wang, Prof. Qingfang Liu
Key Laboratory for Magnetism and Magnetic Materials of the Ministry of Education, Lanzhou University, Lanzhou, 730000, People's Republic of China.

Chengkun Song, Nico Kerber, Jan Rothörl, Yuqing Ge, Klaus Raab, Boris Seng, Maarten A. Brems, Florian Dittrich, Dr. Robert M. Reeve, Dr. Peter Virnau, Prof. Mathias Kläui
Institute of Physics, Johannes Gutenberg-University Mainz, Staudingerweg 7, 55128 Mainz, Germany.
E-mail: klaeui@uni-mainz.de

Nico Kerber, Boris Seng, Dr. Peter Virnau, Prof. Mathias Kläui
Graduate School of Excellence Materials Science in Mainz, 55128 Mainz, Germany.

Boris Seng
Institut Jean Lamour, UMR CNRS 7198, Université de Lorraine, 54506 Vandoeuvre-lès-Nancy, France.

Magnetic skyrmions are topological magnetic structures, which exhibit quasi-particle properties and can show enhanced stability against perturbation from thermal noise. Recently, thermal

Brownian diffusion of these quasi-particles has been found in *continuous films* and applications in unconventional computing have received significant attention, which however require structured elements. Thus, as the next necessary step, we here study skyrmion diffusion in *confined geometries* and find it to be qualitatively different: The diffusion is governed by the interplay between the total number of skyrmions and the structure geometry. In particular, we ascertain the effect of circular and triangular geometrical confinement and find that for triangular geometries the behavior is drastically different for the cases when the number of skyrmions in the element is either commensurate or incommensurate with a symmetric filling of the element. This influence of commensurability is corroborated by simulations of a quasi-particle model.

# 1. Introduction

Magnetic skyrmions are particle-like magnetization whirls with topologically enhanced stability, which have been discovered in a large range of systems, including bulk materials and magnetic films.[1–7] They are found to exist over a wide range of temperatures, with critical temperatures for their disappearance ranging from a few K to far above room temperature depending on the system.[8–14] Owing to their potential small size, dynamics at low driving current, and topological properties, magnetic skyrmions have attracted intense interest in fundamental research and a wide variety of applications have been proposed based on these attributes. Current induced skyrmion motion could be used, for example, in racetrack memory, logic devices, and spin-transfer nano-oscillators.[15–19] An important issue is current induced Joule heating and in general the temperature of the environment. In some materials and for certain applications such as a skyrmion-based racetrack memory device[15] that relies on deterministic dynamics,[20] thermal noise can be a severe problem that can even be a dominant effect determining the skyrmion nucleation, stability and addressability.[21–23] In contrast, non-deterministic thermal dynamics has actually been exploited recently for alternative types of devices that aim at non-conventional logic, such as reservoir computing,[24] and in particular skyrmion-based probabilistic computing[25] and token-based Brownian computing.[26]

Similar to finite temperature Brownian motion for conventional particles,[27] which has been observed in many branches of science, thermally excited skyrmions also exhibit diffusive Brownian motion characteristics, as recently observed experimentally in low-pinning multilayers.[25] The skyrmion diffusion behavior depends, among other things, on the skyrmion spin structure, the system temperature, and applied fields.[25] In particular, an in-plane field induces skyrmion deformations leading to anisotropic diffusion.[28] Furthermore, the diffusion constant was found to increase as a function of temperature and can be controlled by electric fields that tune the magnetic properties.[26] Finally, the influence of the skyrmion winding number on the diffusion has also been studied.[29] While these studies have been carried out on effectively "infinite" systems such as continuous films, where the skyrmion motion on the timescale of the observation is much smaller than the sample dimensions, one would expect that for devices, geometrical confinement effects could drastically alter the dynamics. In particular, the symmetry of the geometry is expected to change the dynamics and is key to the operation of devices where diffusion is employed. Theoretically Schäffer et al. predicted two different time scales of thermal driven motion of ensembles of magnetic skyrmions in a confined geometry.[30] Here, the dynamics depends both on the interaction between skyrmions and edges, as well as on the interaction between the skyrmions themselves. Using micromagnetic and Monte Carlo simulations in a quasi-particle model, Schäffer et al. calculated the possible pattern formations of skyrmions in circular and triangular confined geometries.[30] In particular, they suggested that configurations of skyrmions in triangular confinement with a number of skyrmions that is commensurate with the confinement (in their case 1,3 and 6) are particularly stable. In general, knowledge of the possible skyrmion arrangements in a confined geometry is crucial for using skyrmions as information carriers in data storage, and the dependence of the thermal dynamics on different arrangements and geometries is not clear. In particular, the number of skyrmions in the element compared to the geometrical size is expected to change the relative influence of the skyrmion-edge and skyrmion-skyrmion interactions and an understanding of this is key for the control and eventual employment of thermal skyrmion dynamics in a device. The study of diffusion in extreme confinement conditions is also a necessary requirement for the experimental realization of basic non-conventional computational components, such as C-joins, ratchets, hubs and wires, and the eventual realization of elementary circuits for token-based Brownian computing.[26,28]

In this work, we experimentally probe the dramatic influence of confinement on diffusion, and study the skyrmion diffusion in confined geometries varying both the symmetry of the elements and the number of skyrmions within the element, studying the evolution from a sparse population to a fully lattice-like situation. Even though the diffusion of unconfined skyrmions has been analyzed theoretically,[31] in simulations[32] and recently by experiments,[25] the situation is far less clear in cases of extreme confinement.[30] By tuning the number of skyrmions in a confined geometry and investigating both isotropic (circle) and anisotropic (triangle) structures fabricated from a low pinning Ta/CoFeB/Ta/MgO/Ta multilayer system, we analyze skyrmion stability and confinement effects and reveal a strong dependence of the dynamics on the geometry and skyrmion population. In the triangular geometry, we find that the dynamics varies drastically for numbers of skyrmions that can form a regular order, thus being commensurate with the geometry, as compared to numbers of skyrmions that are incommensurate with the geometry. Corresponding thermal dynamics and mean square displacements (MSDs) are quantitatively identified and corroborated by Brownian dynamics simulations of the Thiele equation applying the same quasi-particle potentials as derived from the micromagnetic simulations in Ref. [30]. While quasi-particle simulations have only been considered for configurations with up to two particles in Ref. [30], here we provide simulations for all the particle numbers under consideration and compare the results directly to experiments. Going beyond the analysis in Ref. [30], we are also able to investigate dynamics in extreme confinement conditions by measuring MSDs both in simulations and experiments, which has not been considered before.

## 2. Results and Discussion

Multilayer stacks of Ta(5)/Co$_{20}$Fe$_{60}$B$_{20}$(0.9)/Ta(0.08)/MgO(2)/Ta(5) were deposited by DC magnetron sputtering, as previously employed (Ref. [25]) and with the thickness of individual layers given in nanometers in parentheses. In this system it has been ascertained that the observed bubbles have a non-trivial topology and thus exhibit skyrmion character.[25] Our sample exhibits perpendicular magnetic anisotropy (PMA). To observe skyrmions and image their Brownian motion, polar magneto-optic Kerr effect (MOKE) microscopy was performed, with a time resolution of 62.5 ms. The sample temperature is set to 341.5 K using a Peltier element so that we can obtain appropriate skyrmion diffusion and stabilize multiple skyrmions in confined geometries. To study confinement effects, we patterned circular and triangular structures from the magnetic material using electron beam lithography (EBL) and Ar ion etching. These particular geometries have been chosen since the circular structure exhibits isotropic circular symmetry whereas the triangular structure possesses a lower symmetry. The diameter of the circle is 16.8 µm and the side length of the equilateral triangle is 26.8 µm.

To explore the thermal diffusion of skyrmions in confinement, we need to develop the necessary tools to study skyrmions in different arrangements with different numbers of skyrmions and determine the interplay between the number of skyrmions and the size of the element. Then we can explore the arrangements of skyrmions depending on the element geometry, where we choose a highly symmetric circular geometry and a lower symmetry triangular geometry. As a first step, we therefore need to nucleate different numbers of skyrmions in our elements. This process is described in the **Supporting Information Figure S1.** Note that the systematic error in the measurement of 0.13 µm dominates in the evaluation of the skyrmion size, whereas the statistical error is only a few nanometers and therefore negligible. To study the Brownian motion under nominally identical conditions, we adjust the out-of-plane (OOP) field to keep the skyrmion diameter $d_s \sim 1.9$ µm for all configurations in the following measurements.

First, we compare the dynamics for different numbers of identical-diameter skyrmions in the geometry with the highest symmetry; the circular geometry. In **Figure 1**, the results for 2, 5 and 8 skyrmions in a disk are depicted. **Figure 1**a shows the time-averaged OOP magnetization component, calculated over 9600 snapshots. The corresponding skyrmion trajectories are shown in **Figure 1**b, where we see that the complex trajectories emerge as a result of the competition between repulsive skyrmion interactions that keep the skyrmions apart and skyrmion-edge repulsion which leads to circular outer contour limits of the skyrmions trajectories close to the boundaries. In particular as the circular geometry exhibits a circular symmetry, the resulting probability distributions for the skyrmion positions are circularly symmetric, and the bright skyrmions form a bright ring-like area of contrast. The contrast along the 360° angular rotation within the disk is a result of the finite measurement time and possible preferred positions due to a non-flat energy landscape. For an increasing number of skyrmions, the repulsive interactions lead to an increase of the radius of the resulting bright ring, until one of the skyrmions is localized in the center (see the example of 8 skyrmions). Finally, we analyze in **Figure 1**c, the OOP magnetization component $m_z(r)$ as a function of the distance from the center to the edge, which is calculated from the integration of the MOKE signals along the angular direction. For 8 skyrmions, in addition to the increase of the radius of the ring, a signal peak appears at $r = 0$ µm as predicted from the micromagnetic simulations in Ref. [30]. From this we find that the arrangements with higher number of skyrmions form a well-ordered ring around the central skyrmions and during the diffusive motion the combination of the repulsion from the edges and each other keeps this relative arrangement of the skyrmions fixed, with simply a collective rotation of the skyrmions in the ring. This rotation in the experiments is less pronounced compared to corresponding simulations[30] due to the finite measurement time combined with the non-flat energy landscape which yields preferential pinning points on the trajectory where the skyrmions tend to stay for a longer time.

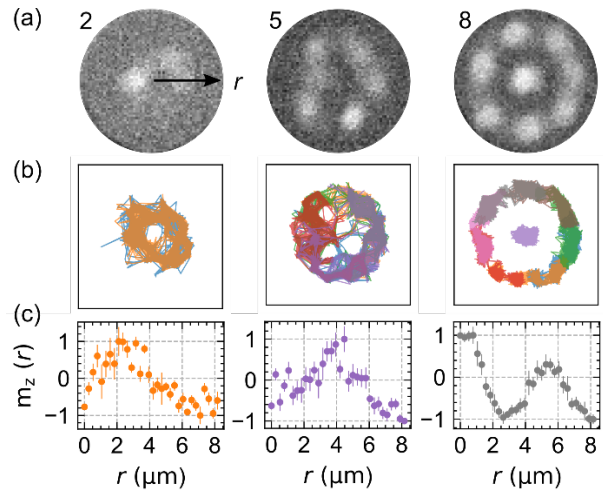

**Figure 1**. (a) Time-averaged out-of-plane magnetization component over 9600 snapshots for the configurations with 2, 5 and 8 skyrmions in the circular geometry. The bright and dark areas represent regions where skyrmions diffuse and unfavorable positions, respectively. (b) Corresponding skyrmion trajectories visualized in different colors and (c) normalized time-averaged signal, averaged over 360° as a function of distance from the center to the edge.

Having established that the filling of a geometry with a certain number of skyrmions impacts the skyrmion arrangements even in a high symmetry structure, we next study a more complex geometry. For the triangular geometry, **Figure 2**a depicts the configurations with an ascending number of skyrmions from 1 to 10 (Top row). As for the circular geometry we set a constant skyrmion size of $d_s \sim 1.9$ μm for all configurations by adjusting the OOP field. The time-averaged frames are shown in the bottom row of **Figure 2**a. The lower symmetry of the triangle results in a more distinct distribution of skyrmion positions, in contrast to the smeared-out trajectories seen for the high symmetry disks. For configurations with 1, 3, 6 and 10 skyrmions, the symmetry displayed for the time-averaged images is in agreement with the initial configurations, which already represent a state that is commensurate with the geometry as predicted by Ref. [30]. Except for the "one skyrmion" configuration, skyrmions arrange themselves into a triangular order, with strong bright contrast at distinct positions due to the confinement resulting from the boundary and skyrmion-skyrmion interactions. The symmetry of the contrast coincides with that of the triangular geometry. For the other configurations, where the number of skyrmions is incommensurate with the triangular geometry skyrmions can diffuse over a comparatively larger area leading to more smeared-out contrast. Time-averaged pictures for these incommensurate states show that in the limited measurement time not all positions have been equally populated, allowing us to ascertain from the dwell times the existence of a non-flat energy landscape.

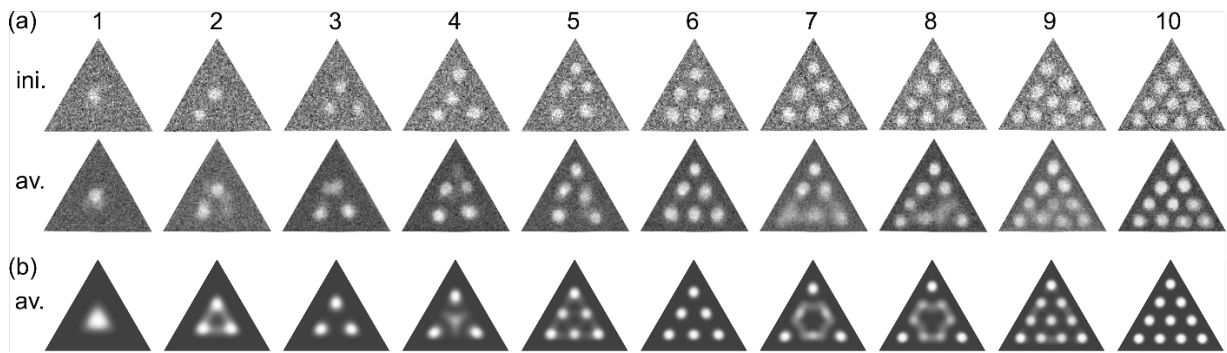

**Figure 2**. (a) Experimental results for skyrmions in a triangular geometry. The top row displays configurations with different initial skyrmion numbers, and the bottom row shows time-averaged

configurations over 9600 Kerr images. (b) Time-averaged trajectories of skyrmions from quasi-particle simulations for configurations of 1 to 10 skyrmions inside the triangle based on positions from 1 million independent frames. Skyrmions are displayed as circles with a radius equal to the skyrmion-skyrmion interaction width $b$.

To corroborate and better understand the intricate dependence of commensurability on skyrmion motion in triangular confinement, we undertake molecular dynamics simulations with a set of overdamped Langevin equations based on the well-established Thiele approach:[33–35]

$$-\gamma v_i = Gz \times v_i + F_i^{SS} + F_i^{SW} + f,$$

where $F^{SS}$ is the skyrmion-skyrmion interaction force, $F^{SW}$ is the interaction force between skyrmions and boundaries and $f$ is thermal white noise. In the Thiele equation, the skyrmionic character enters via the well-established gyrotropic force and magnetic parameters enter passively into G and γ. The damping constant $\gamma$ is set to 1.0, the Magnus force amplitude is set to 0.25 and the timestep is $10^{-5}$. This particular choice for γ and G is motivated by typical experimentally determined skyrmion Hall angles. (Here, we assume Θ~14° [36] with arctan(G/γ)=Θ [37]). The effective coarse-grained potentials for skyrmion-skyrmion and skyrmion-wall interactions applied in this work are based on a micromagnetic description and have been tested and evaluated against micromagnetic simulations in Ref. [30]. Skyrmions are represented by soft disks interacting both with each other and with boundaries via the exponentially decaying potentials:

$$V(r) = ae^{-r/b},$$

with interaction constants $a = 1.246$ eV and $b = 1.176$ nm for skyrmion-skyrmion interaction and $a = 0.211$ eV and $b = 1.343$ nm for skyrmion-boundary interactions, i.e., the latter is slightly softer as expected from the different dipolar interactions. The side length of the triangle is set to 30 nm and the temperature to 15 K in accordance with Schäffer et al.[30] Similar coarse-grained descriptions of skyrmions as quasi-particles can, e.g., be found in Refs. [34,35] and [37] and were also used in our previous work (Refs. [14,25]). We neglect any effects of a non-flat energy landscape, which can be observed in some of the experiments, in our idealized simulations.

As shown in **Figure 2**b, arrangements with triangular symmetry arise for all investigated numbers of skyrmions. In agreement with the experiments, simulations also show states commensurate with a stable triangular geometry for 1, 3, 6 and 10 skyrmions. For other numbers of skyrmions, as in the experiment, no stable triangle can be formed leading to a stronger diffusion of individual skyrmions. In these simulations, skyrmions at the corner of a triangle are mostly stable, while the remaining skyrmions fluctuate between the available positions within the triangular confinement. In the experiments, non-commensurate structures are less symmetric, which can likely be attributed to the non-flat energy landscape that is neglected in the simulations, combined with the finite measurement time. It is worth noting that averaged trajectories from simulations of the quasi-particle model are also in very good agreement with the corresponding micromagnetic simulations from Ref. [30], showing the robustness of the approach.

While qualitatively we see that the commensurability clearly plays a role, we finally explore the diffusion and the impact of the commensurability quantitatively. The (unconstrained) diffusion dynamics of Brownian motion under thermal noise can be expressed as[38,39]

$$< [r_{x,y}(t') - r_{x,y}(t)]^2 > = < (\Delta r_{x,y})^2 > = 4D\Delta t,$$

where the left side expresses the MSD at the positions of the skyrmion center $r_{x,y}$ at $t'$ and $t$, $\Delta t = t' - t$ is the time between two positions and $D$ is the diffusion coefficient. Based on the recorded experimental videos, the MSDs as a function of time are calculated, as shown in **Figure 3**a, where the MSDs are averaged over all skyrmions in the element (MSDs of individual skyrmions can additionally be analyzed to quantify individual skyrmion mobility, as shown exemplary in **Supporting Information Figure S2**.). Two qualitatively different behaviors can be observed: (i) we see that the MSDs of configurations corresponding to commensurate states (1, 3, 6, 10) reach a saturation, and the saturation value of the MSD decreases with ascending numbers of skyrmions. This is in agreement with the concept that a larger number of skyrmions in a commensurate state decreases the space accessible to each skyrmion; (ii) for incommensurate structures, skyrmions cover larger distances and the MSD does not reach a plateau value and thus does not saturate for the observed time scale. Of course, it is expected that eventually all configurations will reach a plateau value given enough time. Analogously to the commensurate states, the MSD is lower for higher numbers of skyrmions. Note that the commensurability effect is generic and not limited to the studied geometry; we also demonstrated this effect for skyrmion diffusion in rectangular wire geometries with different width, see **Supporting Information Figure S4 and Figure S5** and simulation results in **Figure S6**.

**Figure 3**b displays the corresponding simulated values of the short-term MSDs, which have not been calculated previously. Similar to the experiments, the plateau value decreases with increasing numbers of skyrmions for commensurate states. For the incommensurate states, the skyrmions are able to move between different positions and therefore the MSDs are much slower in reaching a saturation value. While the simulation results agree with the experiments qualitatively, there are small differences, which can likely be attributed to a non-flat energy landscape in the experiment and/or our coarse-grained model, particularly for systems with 4 skyrmions, where one skyrmion

is trapped in the center between the other three skyrmions. This trapped position is, however, much less stable than the commensurate states for 3, 6 and 10 skyrmions. Nevertheless, the qualitative behavior is fully reproduced, showing that commensurability effects clearly govern the dynamics in confined geometries. In **Supporting Information (Figure S7 and S8)** we have additionally simulated skyrmions confined to a square. In this case we observe commensurate states for 1, 4 and 9 skyrmions which are compatible with the square geometry. This demonstrates that the occurrence of commensurate states is indeed a generic feature.

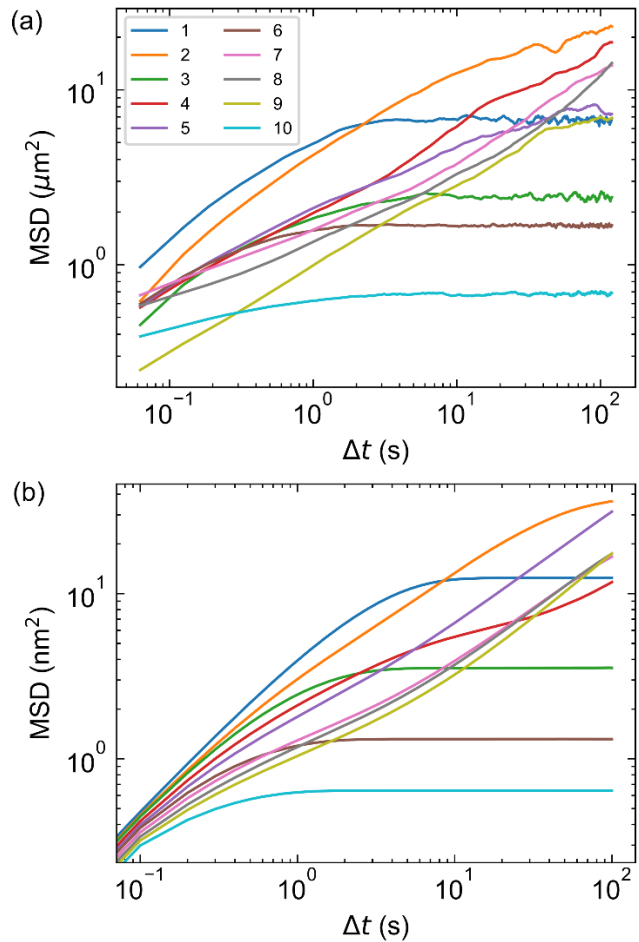

**Figure 3**. MSDs from (a) Experiments and (b) simulations for different skyrmion configurations (represented by different colors) in the triangular geometry as a function of time $\Delta t$.

Note that the computational model system is calibrated by micromagnetic simulations on the nanometer scale,[30] while the experiments actually take place on the scale of micrometers. From a qualitative point of view, however, this does not play a significant role here as the discussed effects are geometric in nature and not really intrinsic to skyrmionic systems. In fact, very similar results for both statics and dynamics can be obtained using generic repulsive potentials and neglecting the skyrmion specific gyrotropic term in the Thiele equation, as demonstrated in **Supporting Information Figure S9 and S10**.

**3. Conclusion**

In conclusion, our investigations of the skyrmion arrangement and diffusion behavior in confined geometries for the magnetic multilayer system $Ta(5)/Co_{20}Fe_{60}B_{20}(0.9)/Ta(0.08)/MgO(2)/Ta(5)$ demonstrate that the intricate interplay between the number of skyrmions in the element and the geometry governs both the arrangement and the diffusion of the skyrmions. In highly symmetric disks we find for larger numbers of skyrmions that multiple skyrmions are arranged in an angularly symmetric manner and the diffusion of skyrmions proceeds via a synchronized rotation along the angular direction. In lower symmetry structures, we discover a commensurability effect in a triangular confinement, where the arrangement and in particular the diffusion qualitatively depend on the commensurability of the skyrmion number and the geometry: By tuning the number of skyrmions to 1, 3, 6 and 10 in the triangle, we find that states commensurate with the geometry are formed and this leads to saturation of the MSDs as a function of time, as predicted by micromagnetic simulations in Ref. [30]. The saturation value for the MSD decreases for higher skyrmion numbers as less space is accessible to individual skyrmions. In contrast, for the incommensurate states, the MSDs exhibit a non-monotonic behavior (as a function of skyrmion numbers) and do not reach a saturation value over the observed timescales. Our molecular dynamics simulations show qualitative agreement with the experimental results, corroborating the basic dependence of the motion of the skyrmion on the skyrmion number for triangular confinement. Our findings confirm experimentally a novel commensurability effect of thermal dynamics of skyrmions in confined geometries and provide a potential method to control diffusive dynamics by tuning the number of skyrmions and the element geometry. In particular our work shows that one needs to consider not only the number of skyrmions when engineering the thermal dynamics but also the commensurability with the geometry, leading to behavior that varies strongly for commensurate and incommensurate fillings of an element. Furthermore, we are able to study for the first time the dynamics of skyrmions in extreme confinement conditions experimentally and with quasi-particle simulations. Our study of these extreme confinement conditions will also be useful as a first necessary step towards the construction of basic computational components in Brownian computing, and may eventually pave the way for their experimental realization. This work clearly demonstrates the challenges for controlled thermal dynamics which vary depending on the filling of a geometry with a commensurate or incommensurate number of skyrmions.

**SUPPORTING INFORMATION**

Supporting Information is available from the Wiley Online Library or from the author.

**ACKNOWLEDGMENTS**

The authors acknowledge funding from TopDyn, SFB TRR 146, SFB TRR 173 Spin+X (project A01 #268565370), and the National Natural Science Foundation of China (Grant No. 51771086). The experimental part of the project was additionally funded by the Deutsche Forschungsgemeinschaft (DFG, German Research Foundation) project No. 403502522 (SPP 2137 Skyrmionics). We acknowledge financial support from the Horizon 2020 Framework Programme of the European Commission under FET-Open Grant No. 863155 (s-Nebula) and Grant No. 856538 (ERC-SyG 3D MAGIC). C.S. acknowledges additional financial support from the China Scholarship Council (Grant No. 201906180096) for a research stay at Johannes Gutenberg University Mainz. N.K., B.S., R.M.R and M.K. gratefully acknowledge financial support by the Graduate School of Excellence Materials Science in Mainz (MAINZ, GSC266).

**AUTHOR CONTRIBUTION**

M.K., P.V. Q. L. and J. W. proposed and supervised the study. N.K. and B.S. patterned the different sample geometries. C.S., N.K and K.R. prepared the measurement set-up and conducted the experiments using the Kerr microscope. C.S. evaluated the experimental data with the help of Y.G.. J.R., Y.G. and M.B. performed the molecular dynamics simulations with the help of F.D.. C.S. drafted the manuscript with the help of N.K., J.R., M.K. and P.V. All the authors commented on the manuscript.

**COMPETING INTERESTS**

The authors declare no competing interests.

**DATA AVAILABILITY STATEMENT**

The data that support the findings of this study are available from the corresponding author upon reasonable request.

**CODE AVAILABILITY STATEMENT**

The code that supports the findings of this study is available from the corresponding author upon reasonable request.